\documentclass[aps,prl,twocolumn,floatfix]{revtex4}

\usepackage{epsfig}
\usepackage{amsmath}

\begin{document}
\preprint{DUKE-TH-02-224}

\title{High energy photons from passage of jets through quark gluon plasma}
\author{Rainer J.~Fries}
\affiliation{Department of Physics, Duke University, 
             Durham, NC 27708-0305}
\author{Berndt M\"uller}
\affiliation{Department of Physics, Duke University, 
             Durham, NC 27708-0305} 
\author{Dinesh K.~Srivastava}
\affiliation{Department of Physics, Duke University, 
             Durham, NC 27708-0305} 
\affiliation{Physics Department, McGill University,
             3600 University Street, Montreal, H3A 2T8, Canada} 
\affiliation{Variable Energy Cyclotron Centre, 
             1/AF Bidhan Nagar, Kolkata 700 064, India}            
\date{\today}
\begin{abstract}
We calculate the production of high energy photons from 
Compton scattering and annihilation of a quark jet passing through
a quark gluon plasma produced in a relativistic heavy ion collision. 
The contributions are large and reflect the momentum
distribution of the jets and the initial conditions of the plasma.
\end{abstract}
\maketitle

Relativistic heavy ion collisions are studied with the aim of 
producing a plasma of quarks and gluons (QGP). Photons are considered to be
an important probe for the investigation of the formation and evolution of
such a plasma due to their weak final-state interactions
\cite{Fei:76}. Once produced they carry the information about 
the conditions of the environment in which they were created,
encoded in their momentum distribution, thus providing a
glimpse deep into the bulk of strongly interacting matter. 
Though most of the measured photons have their origin in the decay of 
hadrons after the QGP phase, 
it has become possible to isolate the direct photons
 produced in such collisions \cite{WA98:00}.

The sources of direct photons considered so far include quark
annihilation, Compton scattering, and bremsstrahlung following the
initial hard scattering of partons of the nuclei \cite{Owens:87}, 
as well as thermal photons from the QGP 
\cite{KapLiSei:91,BNNR:92, RPSSS:96,AGKZ:98,ArnMoYaf:01}
and from hadronic interactions in the hot hadronic gas after the 
hadronization of the plasma \cite{KapLiSei:91,XioShuBro:92}.
The pre-equilibrium production of photons has also been investigated by
several authors \cite{Strickl:94}. Results are available for production
from the entire history of the system \cite{SriSin:01}.

In this letter we study a new source of direct photons originating from
the passage of the produced high energy quark jets through the QGP
(jet-photon conversion). A fast quark passing through the plasma will 
produce photons by Compton scattering with the thermal gluons and
annihilation with the thermal antiquarks. This process is higher 
order in $\alpha_s$ compared with photons from initial hard scatterings, 
but it is not a subleading contribution, since it corresponds to double 
scattering, which is enhanced by the size of the system. For cold 
nuclear matter this effect is encoded in multi-parton matrix elements 
which are enhanced by powers of $A^{1/3}$ \cite{LQS:94}. 
Below, we find that this source is at least comparable in strength 
to the other direct photon sources and even dominates in the range 
$p_\perp \le 6$ GeV for Au+Au collisions at 
the Relativistic Heavy Ion Collider (RHIC).

We also demonstrate that the $p_\perp$-distribution of these photons is 
directly proportional to the momentum distribution of jets at an 
early stage after their production, before they have lost energy on their 
travel through the plasma. Since the measured high-$p_\perp$ hadron 
spectrum is proportional to the spectrum of partons after they have
left the plasma, a comparision of both spectra could provide a 
quantitative determination of the energy loss and help confirm
the mechanism of jet quenching \cite{GyulWang:94,BDMPS:97}.

Furthermore, the photon yield depends on the integrated density of 
the matter traversed by the jets and thus can provide a measurement
of this quantity. We emphasize that our mechanism is distinct from
the back-to-back correlation among direct photons and leading hadrons, 
which was proposed as a direct measurement of the jet energy loss 
in dense matter \cite{WangHS:96}.

The idea of jet-photon conversion is based on the properties of the 
annihilation and Compton cross sections. The kinematics of 
the annihilation of a quark antiquark pair 
$(q + \overline{q} \rightarrow \gamma + g)$
is expressed in terms of the Mandelstam variables
$s= (p_q+p_{\overline{q}})^2$, $t= (p_q-p_{\gamma})^2$ and
$u= (p_{\overline{q}}-p_{\gamma})^2$, 
where the $p_i$ are the four-momenta of the particles.
The differential cross section at Born level for massless partons 
has the form
$  {d\sigma}/{dt} = {8\pi\alpha\alpha_s e_q^2}
  \left( {u}/{t}+ {t}/{u} \right)/{9 s^2} $,
where $e_q$ is the fractional charge of the annihilating quarks.
This implies that the largest  contribution to the production of
photons arises from small values of $t$ or $u$, corresponding to
${\bf p}_\gamma \approx {\bf p}_q$ or ${\bf p}_{\gamma} 
\approx {\bf p}_{\overline{q}}$. 
The process can be visualized as a conversion of one of the 
annihilating quarks to a photon.
This picture immediately suggests that the photons 
from this process provide a direct measurement of the quark (or 
antiquark) momentum.
For the Compton process $(g + q \rightarrow \gamma + q)$
we have
$  {d\sigma}/{dt} = -{\pi\alpha\alpha_s e_q^2}
  \left( {u}/{s}+{s}/{u} \right)/{3 s^2} $ .
The dominant contribution now comes from the region of small $u$, when
${\bf p}_\gamma \approx {\bf p}_q$, corresponding to the conversion
of the quark (or the antiquark) into a photon.

Following \cite{Wong:94,Nadeau:93} we 
approximate the invariant photon differential cross sections 
for the annihilation and Compton processes as
\begin{equation}
  E_\gamma \frac{d\sigma^{\rm (a)}}{d^3p_{\gamma}} \approx
  \sigma^{\rm (a)}(s)
  \frac{1}{2}E_\gamma 
  \left[ \delta(\bf{p}_{\gamma}-\bf{p}_q)+
   \delta(\bf{p}_{\gamma}-\bf{p}_{\overline{q}}) \right ]
\label{app:a} 
\end{equation}
\begin{equation}
 E_\gamma \frac{d\sigma^{\rm (C)}}{d^3p_{\gamma}} \approx
 \sigma^{\rm (C)}(s)
 E_\gamma 
 \delta(\bf{p}_{\gamma}-\bf{p}_q).
\label{app:c} 
\end{equation}
Here $\sigma^{\rm (a)}(s)$  and $\sigma^{\rm (C)}(s)$ are 
the corresponding total cross sections.
Now consider an ensemble of fast quarks passing through a hot medium.
A certain fraction, determined by the total cross section, will undergo 
these annihilation or Compton processes initiated by the medium 
and will emit photons. 
Whenever this happens the energy of the emitted 
photon will reflect the initial energy of the converted quark.

Using (\ref{app:a},\ref{app:c}) the rate of production of 
photons due to annihilation and Compton scattering are given by 
\cite{Wong:94}
\begin{widetext}
\begin{eqnarray}
  E_{\gamma}\frac{dN^{\rm (a)}}{d^4x \, d^3p_{\gamma}}
  &=&\frac{16 E_\gamma}{2(2\pi)^6}
  \sum_{q=1}^{N_f} f_q({\bf p}_\gamma) \int d^3p
  f_{\bar q}({\bf p}) \left[ 1+f_g({\bf p})\right] 
  \sigma^{\rm (a)}(s)
  \frac{\sqrt{s(s-4m^2)}}{2E_\gamma E}
  + (q \leftrightarrow \overline{q}) \>,
  \label{eq:ann}  \\
  E_{\gamma}\frac{dN^{\rm (C)}}{d^4x \, d^3p_{\gamma}}
  &=& \frac{16 E_\gamma}{(2\pi)^6} 
  \sum_{q=1}^{N_f} f_q({\bf p}_\gamma) \int d^3p
  f_g({\bf p})   \left[ 1-f_q({\bf p})\right] 
  \sigma^{\rm (C)}(s)
  \frac{(s-m^2)}{2E E_\gamma } 
  + (q \rightarrow \overline{q}) \>.
  \label{eq:comp}
\end{eqnarray}
\end{widetext}
The $f_i$ are distribution functions for the quarks, antiquarks, and 
gluons. Inserting thermal distributions for the gluons and quarks one 
can obtain an analytical expression for these emission rates for
an equilibrated medium \cite{KapLiSei:91,BNNR:92,Wong:94}.

We propose that the phase-space distribution of the quarks and gluons 
produced in a nuclear collision can be approximately decomposed into 
two components, a thermal component $f_{\rm th}$ characterized by a 
temperature $T$ and a hard component $f_{\rm jet}$ given by hard
scattering of partons and limited to transverse momenta 
$p_{\perp} \gg 1$ GeV/$c$:
$ f({\bf p})=f_{\rm th}({\bf p})+f_{\rm jet}({\bf p})$.
We note that $f_{\rm jet}$ dominates for large momenta, while 
at small momenta $f$ is completely given by the thermal part.

The phase-space distribution for the quark jets propagating through
the QGP is given by the perturbative QCD result for the jet yield
\cite{LinGyu:95}:
\begin{multline}
  f_{\rm jet}({\bf p})=\frac{1}{g_q}\frac{(2\pi)^3}
  {\pi R_\perp^2 \tau p_\perp} 
  \frac{dN_{\rm jet}}{d^2p_\perp dy} \> R(r) \\   \times
  \delta (\eta-y) \Theta(\tau-\tau_i) \Theta (\tau_{\rm max} - \tau) 
  \Theta(R_\perp - r)  \>.
\label{fjet}
\end{multline}
where $g_q=2\times 3$ is the spin and colour degeneracy of the
quarks, $R_\perp$ is the transverse dimension of the system, 
$\tau_i \sim 1/p_\perp$ is the formation time for the jet and $\eta$ is the 
space time rapidity. $R(r)$ is a transverse profile function.
We shall take $\tau_{\rm max}$ as the smaller of the life-time $\tau_f$ of 
the QGP and the time $\tau_d$ taken by the jet produced at position $\bf r$ 
to reach the surface of the plasma. The boost invariant correlation 
between the rapidity $y$ and $\eta$ is assumed here \cite{Bj:83}. 
Similar results 
are obtained when a Gaussian correlation between $y$ and $\eta$, 
characterized by a width $\Delta_p \sim 2/p_\perp \cosh y$ is
assumed \cite{LinGyu:95}.

We calculate the jet production and the direct production of photons 
in lowest order pQCD \cite{Owens:87}.
We have used the CTEQ5L parton distributions \cite{cteq5} and EKS98 
nuclear modifications \cite{EKS98} with the scale set to $p_\perp$. 
We further use a $K$-factor of 2.5 in the jet production to account for 
higher order corrections, cf.\ \cite{EskTuo:00}. 
We neglect the weak dependence of $K$ on $p_\perp$ and $\sqrt{S}$ 
in this study \cite{BFLPZ:00}. Our results for the number of quarks, 
antiquarks and gluons at $y=0$ can be parameterized as 
\begin{equation}
  \frac{d N^{\rm jet}}{d^2 p_\perp \, d y} \> \bigg|_{y=0} =
  T_{AA} \> \frac{d \sigma^{\rm jet}}{d^2p_\perp \, dy} \> \bigg|_{y=0} 
  = K \frac{a}{(1+p_\perp/b)^{c}} \> \, .
  \label{eq:para}
\end{equation}
Here $T_{AA}=9A^2/8\pi R_\perp^2$ is the nuclear thickness for a
head-on collision. Numerical values for the parameters $a$, $b$ and 
$c$ are listed in Table \ref{tab:minijets}.

The integral over $\bf p$ in Eqs.~(\ref{eq:ann}) and (\ref{eq:comp}) is 
dominated by small momenta. We can therefore drop the jet part in the 
decomposition of the distributions $f({\bf p})$ in the integrands and 
approximate them by the thermal part. 
After performing the integrals the results for annihilation and Compton 
scattering are identical and read \cite{Wong:94}
\begin{multline}
  E_\gamma \frac{dN_\gamma^{(\rm a)}
  }{d^3p_\gamma \, d^4 x} =  E_\gamma \frac{dN_\gamma^{(\rm C)} 
  }{d^3p_\gamma \, d^4 x}
  =\frac{\alpha \alpha_s}{8\pi^2}
  \sum_{f=1}^{N_f} \left(\frac{e_{q_f}}{e}\right)^2 \\  \times
  \left[ f_q (p_\gamma) +f_{\overline{q}}
  (p_\gamma)\right]
  T^2 \left[ \ln \left\{ \frac{4E_\gamma T}{m^2} \right\} +C
  \right]
  \label{eq:annil}
\end{multline}
with $C=-1.916$. We have included the three lightest quark 
flavours so that $\sum_f e_{q_f}^2 / e^2 = 2/3$.

Once again we apply the decomposition of the distributions to 
(\ref{eq:annil}). Inserting the thermal part we recover
the known rate of emission of high energy photons from interactions 
within the plasma. Since we are dealing with (nearly) massless partons 
we have to address the infrared divergence in (\ref{eq:annil}) in the
limit $m\to 0$. It has been shown \cite{KapLiSei:91,BNNR:92} that the 
divergence can be eliminated by including higher-order effects from the
interaction of the quarks in the thermal medium with the result that 
$m^2$ in the argument of the logarithm must be replaced with 
$2 m_{\rm th}^2= g^2 T^2 /3$.
We assume that these considerations remain valid for the emission
of photons even when one of the partons is from a jet while the other one 
is from a thermal medium. This assumption remains to be verified.
We add that  the thermal photon production from the plasma has recently 
been calculated up to two loops \cite{AGKZ:98} and to complete leading 
order \cite{ArnMoYaf:01}, and is about a factor of two larger than the 
lowest order results given here. It will be of considerable
interest to extend our present work to account for these effects.

\begin{table}[tb]
\begin{tabular}{||c|c|c|c|c||}
  \hline\hline
   &  & 
  \begin{minipage}{2cm}\begin{center} {$a \quad [1/{\rm GeV}^2]$}
  \end{center}
  \end{minipage} & 
  \begin{minipage}{1.8cm}\begin{center} {$b \quad [{\rm GeV}]$ }
  \end{center}\end{minipage} & 
  \begin{minipage}{1.2cm}\begin{center}{$c$}\end{center}
  \end{minipage}   \\ \hline\hline
  RHIC & q & 5.0$\times 10^2$ & 1.6 & 7.9 \\ \cline{2-5}
  & $\overline{\rm q}$ & 1.3$\times 10^2$ & 1.9 & 8.9 
  \\ \hline\hline
  \hspace{0.7em}LHC\hspace{0.7em} & \hspace{0.5em}q\hspace{0.5em} 
  & 1.4$\times 10^4$ & 0.61 & 5.3  \\ \cline{2-5}
  & $\overline{\rm q}$ & 1.4$\times 10^5$ & 0.32 & 5.2 
  \\ \hline\hline
\end{tabular}
  \caption{Parameters for the  minijet distribution 
  $dN/d^2 p_\perp \, dy$ given in Eq.~(\ref{eq:para}) at $y=0$ for Au+Au at 
  $\sqrt{S_{\rm NN}}=200$ GeV (RHIC) and for Pb+Pb at $\sqrt{S_{\rm NN}}=5.5$
  TeV (LHC). Numbers for quarks and antiquarks are mean values for the three 
  lightest flavours:
  q=(u+d+s)/3, $\overline{\rm q}$=($\overline{\rm u}$+$\overline{\rm d}$+
  $\overline{\rm s}$)/3. $K=2.5$. The range of validity is 
  2 GeV/$c < p_\perp < 20$ GeV/$c$.}
  \label{tab:minijets}
\end{table}

Now the rate of high energy photon emission from the passage of quark 
jets through the QGP is obtained by substituting the jet contributions
for $f_q$ and $f_{\overline{q}}$ in (\ref{eq:annil}). In order to 
calculate the photon yields we specify our model for the medium. 
Here we are interested in the emission of photons from the QGP phase alone.
We assume that a thermally and chemically equilibrated plasma
is produced in the collision at time $\tau_0$ with temperature $T_0$, 
and we ignore the transverse expansion of the plasma. Further
assuming an isentropic longitudinal expansion \cite{Bj:83}, 
we can relate $T_0^3 \tau_0$ to the observed particle rapidity density 
$dN/dy$. We take $dN/dy \approx 1260$, based on the charged particle 
pseudorapidity density measured by the PHOBOS experiment 
\cite{Phobos:02} for central collisions of Au nuclei at 
$\sqrt{S_{NN}}=$ 200 GeV. For central collision of Pb nuclei at LHC 
energies we use  $dN/dy=5625$ estimated by \cite{KaMcLSri:92}. Imposing 
a rapid thermalization limited by $\tau_0 \sim 1/3 T_0$ 
\cite{KaMcLSri:92} we fix our initial conditions to be 
$T_0=446$ MeV and $\tau_0 = 0.147$ fm/$c$ for RHIC and 
$T_0=897$ MeV and $\tau_0 = 0.073$ fm/$c$ for LHC. 

We model the nuclei as uniform spheres and assume a transverse 
profile for the initial temperature as 
$T(r) = T_0 [2 (1-r^2/R_\perp^2)]^{1/4}$ where $R_\perp= 1.2 A^{1/3}$ fm. 
We use the same profile $R(r)=2(1-r^2/R_\perp^2)$ for the jet production
in (\ref{fjet}) while performing the space-time integration 
$d^4x=\tau d\tau \, r dr \, d\eta \, d\phi$.
The limits of the $\tau$-integration are $[\tau_0,\tau_f]$ 
where $\tau_f$ is fixed through the relation $T^3\tau=$ const, 
when the temperature reaches $T_f=160$ MeV. 

We also give results for the competing photon processes, the direct 
production in primary hard annihilation or Compton processes and the 
production via bremsstrahlung from a produced jet parton.  
For both cases the leading order results can again be found in
\cite{Owens:87}. No $K$-factor is used for direct photon production.
Our results here obey the scaling for single photons suggested 
in \cite{Sri:01} from empirical considerations. 

\begin{figure}[tb]
  \begin{center}  
  \epsfig{file=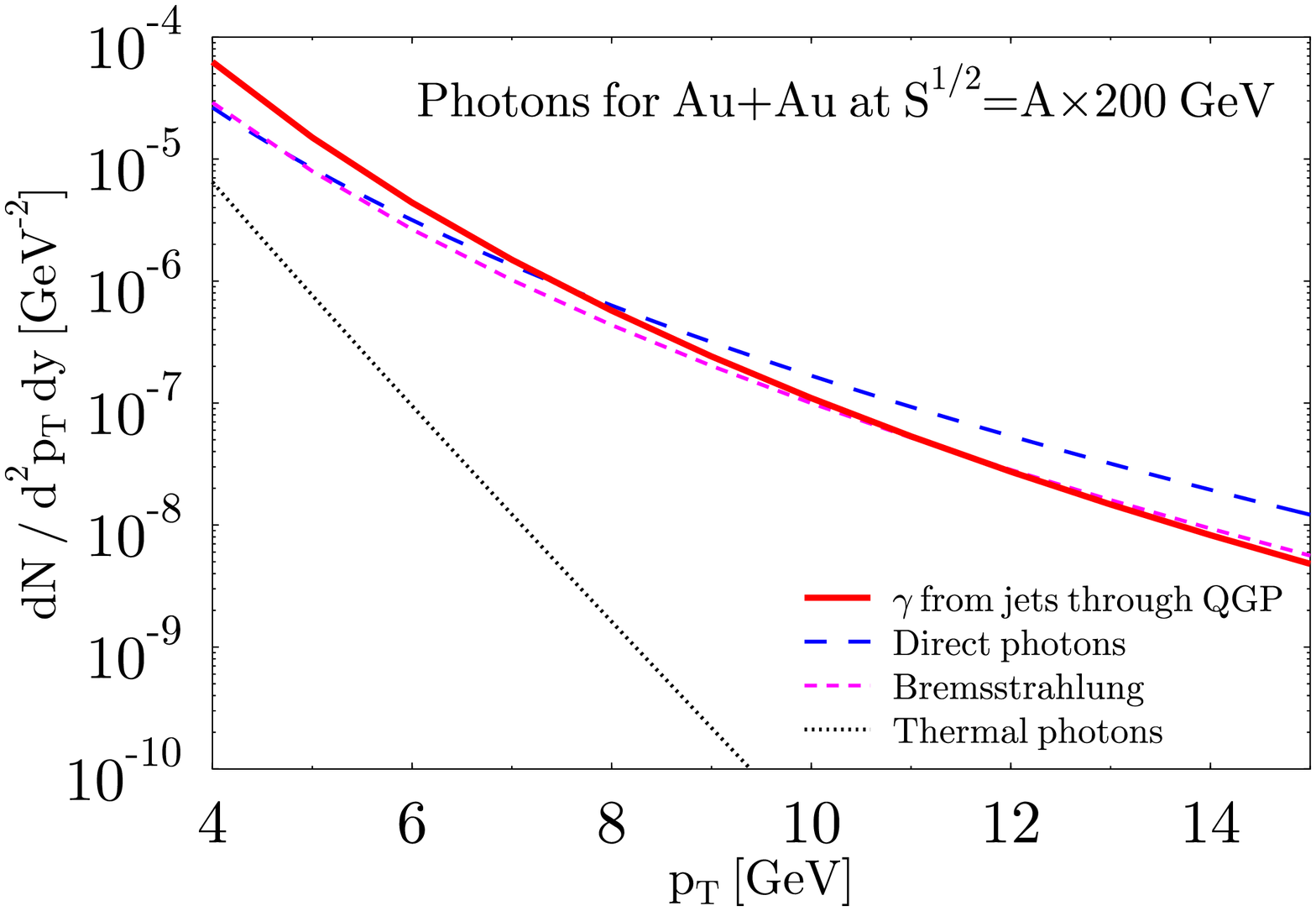,width=7.6cm}
  \caption{Spectrum $dN/ d^2p_\perp dy$ of photons at $y=0$ for central
   collision of gold nuclei at $\sqrt{S_{NN}}=200$ GeV at RHIC. 
   We show the photons
   from jets interacting with the medium (solid line), direct hard photons
   (long dashed), bremsstrahlung photons (short dashed) and 
   thermal photons (dotted).}
  \label{fig:photonrhic}
  \end{center}
\end{figure}

\begin{figure}[tb]
  \begin{center}  
  \epsfig{file=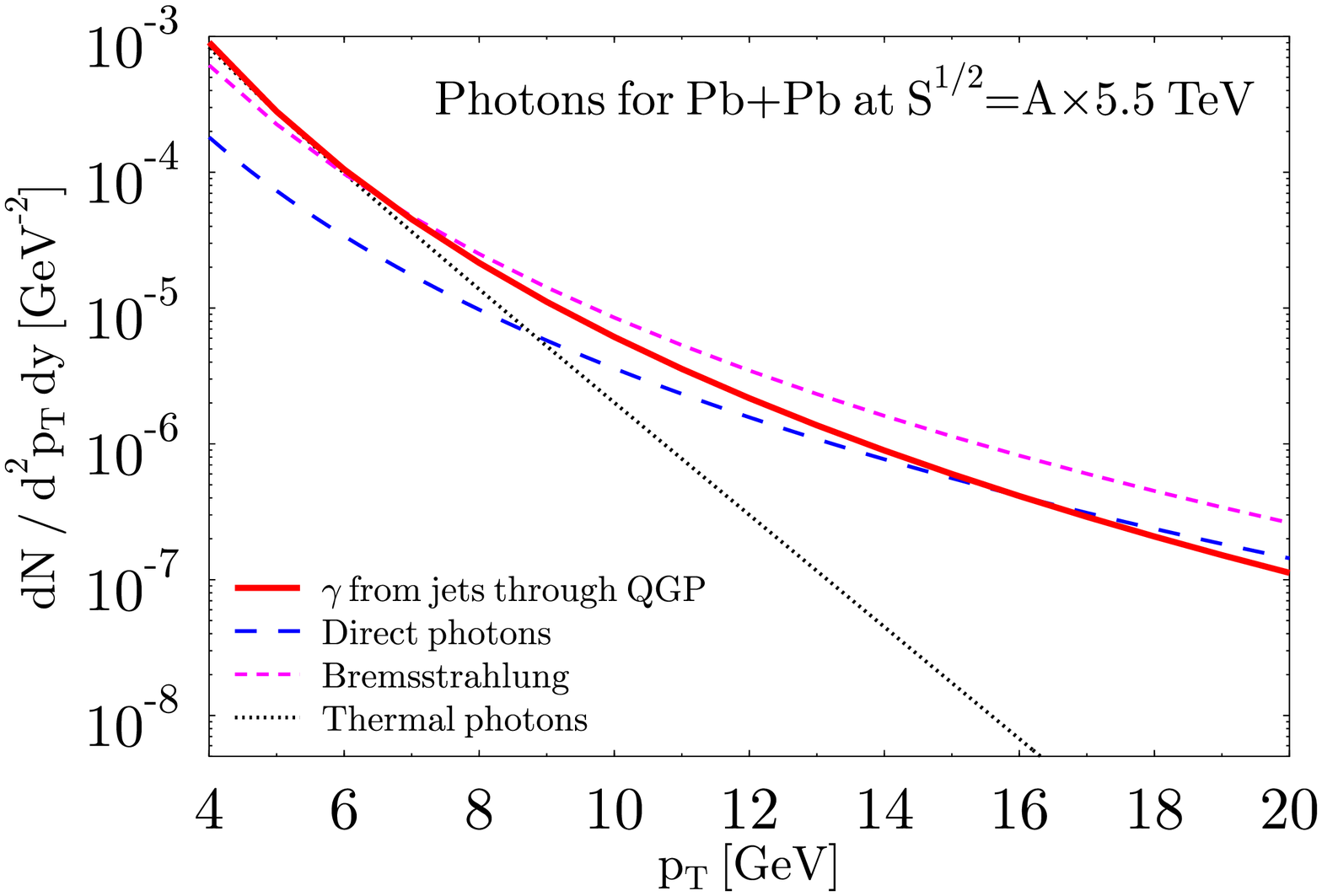,width=7.6cm}
  \caption{The same as Fig.~\ref{fig:photonrhic} for central collision
  of lead nuclei at $\sqrt{S_{NN}}=5.5$ TeV at LHC.}
  \label{fig:photonlhc}
  \end{center}
\end{figure}

In Fig.~\ref{fig:photonrhic} we plot the results for thermal photons, 
direct photons due to primary processes, bremsstrahlung photons and the 
photons coming from jets passing though the QGP in central collision
of gold nuclei at RHIC energies. The corresponding results for LHC energies 
are shown in Fig.~\ref{fig:photonlhc}. We see that the quark jets passing 
through the QGP give rise to a large yield of high energy photons. 
This contribution should be absent in $pp$ collisions. 
For RHIC this contribution is the dominant source of photons
up to $p_\perp \approx 6$ GeV. Note that the jet-to-photon 
conversion falls more rapidly with $p_\perp$ than the direct 
photon yiield, similar to a higher twist correction. 

A suppression of the bremsstrahlung contribution due to the multiple
scattering suffered by the fragmenting partons~\cite{ina} will further 
enhance the importance of the jet-photon conversion process.
The effects of a possible chemical non-equilibrium  may be estimated 
as follows. A reduced quark abundance will enter at least linearly into the
thermal rate but can reduce the jet-photon conversion by at 
most a factor of two, as the Compton contribution will remain unaffected.
Therefore the thermal rate will be more strongly suppressed than the 
jet-photon conversion. Similarly, the jet-photon conversion is only
moderately sensitive to the initial conditions and the transit time 
of the jet in the medium. We have found that increasing $\tau_0$ to 
1 fm/$c$, keeping $T_0^3 \tau_0$ fixed, reduces our results for RHIC 
only by about 40 percent.

One may ask how our results will be modified by the energy loss experienced
by the quark jets passing through the QGP. This effect will be relatively
small for two reasons. First, we are concerned here with quark jets which
lose energy at less than half the rate as gluons, which form the dominant
fraction of jet events. Second, as the energy loss scales as the square of
the distance traveled through the medium \cite{BDMPS:97}, the average 
energy loss at the moment of jet-photon conversion is only one-third of 
the total normal energy loss. 
This argument implies that, while the high-$p_\perp$ hadron 
spectra encode information about the energy loss in the plasma, the 
photons from medium induced emission, on average, carry information about 
jets at an earlier stage. Obviously, the comparison of the high-$p_\perp$ 
hadron and direct photon spectra can shed light on the evolution of the 
jet spectrum. We believe that this could be a useful method for the 
analysis of jet quenching and medium effects on hard particles in the QGP.
More details of the calculation together with a careful analysis of the 
assumptions made here will be presented in a forthcoming paper.

In summary, we have discussed a novel mechanism for the production of high 
energy photons in relativistic heavy ion collisions. The photons arise
from the passage of quark jets through the medium, when a quark is 
converted into a photon due to Compton scattering from a gluon 
or annihilation by an antiquark. Our calculations indicate that this 
mechanism is the leading source of directly produced photons at RHIC 
in the region $p_\perp \le 6$ GeV/$c$. The spectrum of the resulting 
photons reflects the initial distribution of the hard scattered quarks 
and is sensitive to the initial conditions and the traversal time of the
jet in the plasma.

\begin{acknowledgments}  
This work was supported in part by DOE grants DE-FG02-96ER40945 and
DE-AC02-98CH10886 and the Natural Sciences and Engineering Research
Council of Canada.  RJF is supported by the Feodor Lynen program of 
the Alexander von Humboldt Foundation. 
\end{acknowledgments}

\end{document}